\documentclass[reqno,11pt]{amsart}
\usepackage[utf8]{inputenc}
\usepackage{enumitem}
\usepackage[linktocpage]{hyperref}
\usepackage{graphicx}
\usepackage{amscd}
\usepackage{slashed}
\usepackage{amssymb}
\usepackage{mathtools} 
\usepackage[off]{auto-pst-pdf} 
\usepackage{epsfig}
\usepackage{pst-grad} 
\usepackage{pst-plot} 
\usepackage[space]{grffile} 
\usepackage{etoolbox} 
\makeatletter 
\patchcmd\Gread@eps{\@inputcheck#1 }{\@inputcheck"#1"\relax}{}{}
\makeatother

\usepackage[mathscr]{eucal}
\numberwithin{equation}{section}
\allowdisplaybreaks[1]

\title[The Fock Space Dynamics of CFS: Non-Abelian Gauge Fields]
{The Fock Space Dynamics of Causal Fermion Systems:
Non-Abelian Gauge Fields}

\author[C.\ Dappiaggi]{Claudio Dappiaggi}
\address{Dipartimento di Fisica \\ Universit{\`a} degli Studi di Pavia and INFN, Sezione di Pavia \\ Via Bassi, 6 --  I-27100 Pavia \\ Italy}
\email{claudio.dappiaggi@unipv.it}

\author[F.\ Finster]{Felix Finster}
\address{Fakult\"at f\"ur Mathematik \\ Universit\"at Regensburg \\ D-93040 Regensburg \\ Germany}
\email{finster@ur.de}

\author[N.\ Kamran]{Niky Kamran}
\address{Department of Mathematics and Statistics \\ McGill University \\ Montr{\'e}al \\ Canada}
\email{nkamran@math.mcgill.ca}

\author[M.\ Reintjes]{Moritz Reintjes \\ July / October 2026}
\address{Department of Mathematics \\ City University of Hong Kong \\ SAR Hong Kong}
\email{moritzreintjes@gmail.com}

\newtheorem{Def}{Definition}[section]
\newtheorem{Thm}[Def]{Theorem}

\newcommand{\Thanks}{\vspace*{.5em} \noindent \thanks}
\newcommand{\beq}{\begin{equation}}
\newcommand{\eeq}{\end{equation}}
\newcommand{\Proof}{\begin{proof}}
\newcommand{\QED}{\end{proof} \noindent}

\newcommand{\ket}{\mathclose{>}}

\newcommand{\C}{\mathbb{C}}
\newcommand{\R}{\mathbb{R}}

\newcommand{\Pdd}{\mbox{$\partial$ \hspace{-1.2 em} $/$}}

\newcommand{\U}{{\rm{U}}}

\newcommand{\G}{\mathscr{G}}
\newcommand{\g}{\mathfrak{g}}

\newcommand{\bep}{\begin{pmatrix}}
\newcommand{\enp}{\end{pmatrix}}

\renewcommand{\O}{\mathscr{O}}

\newcommand{\Dir}{{\mathcal{D}}}

\newcommand{\B}{{\mathscr{B}}}
\renewcommand{\O}{{\mathscr{O}}}

\newcommand{\scrU}{{\mathscr{U}}}

\newcommand{\macro}{{\mathrm{\tiny{macro}}}}

\newcommand{\starD}{\triangleright}
\newcommand{\hol}{\text{\rm{hol}}}
\newcommand{\bbra}{\mathopen{\ll}}
\newcommand{\kket}{\mathclose{\gg}}

\DeclareFontFamily{OT1}{rsfso}{}
\DeclareFontShape{OT1}{rsfso}{m}{n}{ <-7> rsfso5 <7-10> rsfso7 <10-> rsfso10}{}
\DeclareMathAlphabet{\myscr}{OT1}{rsfso}{m}{n}

\DeclareMathOperator{\Texp}{Texp}

\newcommand{\bitem}{\begin{itemize}[leftmargin=2em]}
\newcommand{\eitem}{\end{itemize}}

\begin{document}

\maketitle

\begin{abstract}
A limiting case is worked out in which the causal action principle for causal fermion systems describing Minkowski space gives rise to the linear Fock space dynamics of perturbative quantum field theory, including non-abelian gauge fields and Dirac fields.
\end{abstract}

\tableofcontents

\section{Introduction} \label{secintro}
In the recent series of papers~\cite{fockbosonic, fockfermionic, fockentangle, fockdynamics} it was show that, in a well-defined limiting case, the causal action principle gives rise to perturbative quantum field theory (pQFT) in Minkowski space. However, 
so far this connection has been made only for abelian gauge fields,
giving quantum electrodynamics coupled to the Dirac field.
In the present paper, we extend the methods and results to
{\em{non-abelian gauge fields}}. This concludes the research program
of recovering the interactions of the standard model of elementary particles
on the level of second-quantized bosonic fields
(for classical bosonic fields, this connection was already established
in~\cite{cfs}). The survey paper~\cite{qftlimit} provides a
detailed and self-contained review of this research program.

In order to avoid excessive overlap, we keep the presentation short
and refer for the general context and more details to the just-mentioned
research papers and the review~\cite{qftlimit}. We focus on the key questions
which arise when extending our previous analysis to non-abelian gauge fields:
\bitem
\item[(1)] How do the bosonic vertices corresponding to the nonlinearity
in the Yang-Mills equation (like the gluon-gluon vertices) come about?
\item [(2)] How is the local gauge freedom handled?
\eitem
In order to tackle the first question, it no longer suffices to treat the bosonic
field as a linear stochastic background field. Instead, the non-linear Yang-Mills
equations must be included in the dynamical equations.
The interesting and important
question arises how to formulate the Yang-Mills equations for the
stochastic ensemble of bosonic fields as introduced in~\cite{fockdynamics}.
Here the Euler-Lagrange (EL) equations of the causal action principle provide an answer,
because they imply that the Yang-Mills equations must be satisfied as operator
equations for the field operators as derived in~\cite{fockdynamics} from the concept of
holographic mixing. This will be worked out in detail in Section~\ref{secbackground}.
Combining this result with a gauge-fixing procedure, we can proceed
similar as in~\cite{fockdynamics} with a formal perturbation expansion
(Section~\ref{secdyson}). We prove that, in this limiting case, we obtain agreement
with the usual perturbation expansion of pQFT in terms of Feynman diagrams.
In Section~\ref{secBRST}, it is outlined how this perturbation expansion
can be treated in a gauge-covariant way using the BRST formalism.

The paper is organized as follows.
Section~\ref{secprelim} provides the necessary preliminaries.
In Section~\ref{secYMholo} the Yang-Mills equation with holographic mixing
is introduced.
In Section~\ref{secbackground} the gauge freedom is discussed,
and a gauge-fixing procedure is introduced. Moreover, we specify
the covariance of the stochastic background field.
In Section~\ref{secdyson} the perturbation expansion is performed on a formal level,
giving all the Feynman diagrams of pQFT.
In Section~\ref{secBRST}, it is discussed how this formal perturbation series
can be given a mathematical meaning following the familiar path using
renormalization in the BRST formalism.

\section{Preliminaries} \label{secprelim}
As already mentioned, we keep the presentation short and focus on those
aspects needed for our constructions. A more detailed introduction
can be found in~\cite{qftlimit} or the previous papers~\cite{fockfermionic,
fockdynamics}.

\subsection{The Dirac Equation with Nonlocal Stochastic Potential}
Causal fermion systems allow for the description of spacetimes
which, on a microscopic length scale, typically thought of as the Planck
scale, can have a structure which is different from the usual spacetime
continuum. Following the constructions and results in~\cite{nonlocal}, the 
microscopic dynamics may be described by a Dirac equation
involving a nonlocal stochastic potential. We will take the following
results from~\cite{nonlocal} as the starting point:
\bitem
\item[(i)] The dynamics of the wave functions in a causal fermion system describing Min\-kowski space
can be described by a Dirac equation of the form
\beq \label{dirnonloc}
\big( i \Pdd + \B - m \big) \psi = 0 \:,
\eeq
where~$\B$ is a {\em{nonlocal}} potential, i.e.\ an integral operator of the form
\[ 
\big( \B \psi \big)(x) = \int_M \B(x,y)\, \psi(y)\: d^4y \]
involving an integral kernel which is symmetric in the sense that
\beq \label{Bsymm}
\B(x,y)^* = \B(x,y) \:.
\eeq

The form of the kernel~$\B(x,y)$ is determined by the EL equations of
the causal action principle. More specifically, this kernel is nonlocal on a
scale~$\ell_{\min}$ lying between
the length scale~$\varepsilon$ of the ultraviolet (UV) regularization (which can be thought of as the Planck scale)
and the length scale~$\ell_\macro$ of macroscopic physics (which can be thought of as the
Compton scale),
\beq \label{ellmindef}
\varepsilon \ll \ell_{\min} \ll \ell_\macro \:.
\eeq
In simple terms, $\ell_{\min}$ can be regarded as the minimal length scale on which the
analysis on the light cone and the resulting formalism of the continuum limit
as developed in~\cite{pfp, cfs} apply.

\item[(ii)] The nonlocal potential~$\B$ is composed of a collection of vector potentials~$A_a$
with~$a  \in \{1,\ldots N\}$. More precisely,
\beq \label{hatBjdef}
\B(x,y) = \sum_{a=1}^N \slashed{A}_a \Big( \frac{x+y}{2} \Big) \:L_a(y-x) \:,
\eeq
where the~$L_a$ are fixed complex-valued kernels.
The number~$N$ of these fields is very large and scales like
\[ 
N \simeq \frac{\ell_{\min}}{\varepsilon} \:. \]
The EL equations of the causal action principle imply that, in a suitable gauge, all
the potentials~$A_a$ must satisfy the scalar wave equation.
\eitem
For a motivation and a more detailed discussion of the nonlocal potential
we refer to~\cite[Section~2.5]{fockdynamics} and the introduction in~\cite{nonlocal}.

\subsection{Implementing Holographic Mixing}
One of the main findings in~\cite{fockdynamics}
(based on related considerations in~\cite{fockentangle}) is that, for the description of second-quantized bosonic fields, one should take into account dephasing effects.
This is modelled mathematically by potentials in the Dirac equation which
are again nonlocal but, compared to~\eqref{hatBjdef}, also include phase factors.
More precisely, we consider the ansatz~\cite[eq.~(4.18)]{fockentangle}
\beq \label{Bansatz}
\B = \sum_{a,b,c=1}^N e^{i \Lambda_a} \:\big( \slashed{A}_{c} \starD L^c_{a,b} \big)\: e^{-i \Lambda_b} \:,
\eeq
where the triangle stands for pointwise multiplication of the kernels according to
\[ \big( \slashed{A} \starD L \big)(x,y) := \slashed{A} \Big( \frac{x+y}{2} \Big)\:
L(y-x) \]
(we do not make a distinction between upper and lower indices~$a,b,c \in \{1,\ldots, N\}$;
the placement simply is for notational convenience).
Compared to~\eqref{hatBjdef} we now have~$N^3$ complex-valued kernels~$L^c_{a,b}$.
Moreover, the $\Lambda_a$ are real-valued functions which change on small scales.
Accordingly, the corresponding phase factors~$e^{i \Lambda_a}$ oscillate
rapidly on small scales. More precisely, we denote the corresponding length scale
by~$\ell_\Lambda$ and assume that it is much larger than the scale~$\ell_{\min}$,
do not oscillate on the length scale~$\ell_{\min}$ of the nonlocality, i.e.\ if
\beq \label{Lamscale}
\big| D^p \Lambda_a \big| \eqsim \frac{1}{\ell_\Lambda^p}\: |\Lambda_a| \qquad \text{with}
\qquad \ell_\Lambda \gg \ell_{\min} \:.
\eeq
When taking operator products of the nonlocal potentials,
we get integrals over these rapidly oscillating phase factors.
As a consequence, we get contributions only if
all phase factors drop out (up to small error terms as worked out and quantified
in~\cite[Section~4.5]{fockdynamics} with a stationary phase analysis).
These findings can be summarized in the calculation rule
\beq \label{dephaserule}
e^{-i \Lambda_b}\: e^{i \Lambda_c} \approx \delta_{bc} \:,
\eeq
to be understood that integrals of these expression are small,
where~``$\approx$'' means up to small error terms with the scaling
\[ \times \Big( 1 + \O \Big( \frac{\ell_\Lambda}{\ell_\macro} \Big) \Big) \]
(for details see~\cite[Section~4.5]{fockdynamics}).
We refer to the dephasing effect leading to the calculation rule~\eqref{dephaserule}
as {\em{holographic mixing}} (a detailed motivation and discussion can be found
in~\cite[Section~4.1]{fockdynamics}). This name is inspired by
the similarity to a hologram in which several pictures are stored, each of which becomes visible only
when looking at the hologram in the corresponding coherent light
(but it is unrelated to holography in string theory).

When taking powers of the nonlocal potential, this rule can be implemented
most conveniently with a matrix notation. To this end, we set
\beq \label{notmatrix}
\mathbf{L}^c := (L^c_{a,b})_{a,b=1,\ldots, N} \:,
\eeq
and introduce the $N \times N$-matrix-valued potential
\[ \mathbf{B} = \sum_{c=1}^N \big( \slashed{A}_{c} \starD \mathbf{L}^c \big) \:. \]
Then
\beq \label{Bprodapprox}
\B \B \approx \sum_{a,b=1}^N e^{i \Lambda_a} \:
(\mathbf{B}\, \mathbf{B})^a_b \: e^{-i \Lambda_b}
\eeq
with
\[ \mathbf{B}\, \mathbf{B} =
\sum_{c,d=1}^N \big( \slashed{A}_{c} \starD \mathbf{L}^c \big)
\big( \slashed{A}_{d} \starD \mathbf{L}^d \big) \qquad \text{with} \qquad
\big(\mathbf{L}^c\, \mathbf{L}^d \big)^a_b := \sum_{e=1}^N L^c_{ae}  L^d_{eb} \:. \]

In~\eqref{Bprodapprox}, the indices~$a$ and~$b$ of the matrix product determine
the phase factors~$e^{i \Lambda_a}$ and~$e^{-i \Lambda_b}$.
With this in mind, we can simplify the notation further by omitting the phase factors at the
very left and right. This leads us to the formalism where the Dirac operator takes the form
\beq \label{dirhol}
\Dir^\hol = i \Pdd + \mathbf{B} \:.
\eeq
By suitable inserting phase factors into the Green's operators
(for details see~\cite[Section~5.3]{fockdynamics}), it is possible to write
the perturbation expansion for Dirac wave functions in the familiar form
of the form (see~\cite[eq.~(5.14)]{fockdynamics})
\[ 
| \tilde{\psi} \ket \approx \sum_{n=0}^{\infty} \big( -s_m\, \mathbf{B}
\big)^n \,|\psi \ket \:, \]
where~$s_m$ are the Dirac Green's operators. This {\em{holographic perturbation expansion}}
differs from the standard perturbation expansion for Dirac wave functions
in an external potential only by the fact that the potentials are $N \times N$-matrix-valued.
The components of these matrices describe a mixing of the different holographic components.

\section{The Yang-Mills Equation with Holographic Mixing} \label{secYMholo}
The potentials in~\eqref{Bansatz} can be thought of a family of
abelian gauge potentials coupling to the Dirac equation in a non-local
way. In this section, we generalize this setting to non-abelian gauge
fields.

\subsection{The Non-Local Yang-Mills Potential with Holographic Mixing}
We let~$\G$ be a (finite dimensional) Lie group.
We denote its Lie algebra by~$\g := T_e \G$.
We choose a basis~$(e_\alpha)_{\alpha=1,\ldots, L}$ of~$\g$ and write the commutation relations as
\[ [e_\alpha, e_\beta] = \sum_{\gamma=1}^L f_{\alpha \beta \gamma}\, e_\gamma \]
with structure constants~$f_{\alpha \beta \gamma}$. Given a Yang-Mills potential~$A$,
we introduce the corresponding gauge-covariant derivative as usual by
\[ D^A_j = \partial_j - i A_j \:. \]
Its curvature and current are given by
\begin{align}
F^A_{jk} &= i [D^A_j, D^A_k] = (dA)_{jk} - i [A_j, A_k] \\
F^A &= dA - i A \wedge A \\
J^A &= *(D * F^A) = *(d * F^A) - i *(A * F^A) \notag \\
&= *(d * d A) -i *(d * (A \wedge A))- i *(A * d A) + *(A * (A \wedge A)) \:, \label{JYM}
\end{align}
where~$* : \Omega^k(M) \rightarrow \Omega^{n-k}(M)$ is the usual Hodge star.

In order to build the Yang-Mills potentials into the Dirac equation,
we simply replace the potentials~$A_a$
in~\eqref{hatBjdef} by potentials taking values in the Lie algebra, i.e.\
\[ A_{a,j}(x) = \sum_{\alpha=1}^L A^\alpha_{a,j}(x) \: e_\alpha \in \g \:. \]
Correspondingly, we replace the matrix potential $\mathbf{B}$ in the
holographic Dirac equation~\eqref{dirhol} by
\beq \label{Bdef}
\mathbf{B} = \sum_{\alpha=1}^L
\sum_{c=1}^N \slashed{A}^\alpha_{c} \starD \mathbf{L}^c \: e_\alpha \:.
\eeq

We again want to describe the multitude of potentials~$A^\alpha_{a,j}$ stochastically.
However, before we can do this, we need to specify the dynamical equations
for the family of Yang-Mills potentials.

\subsection{The Field Equations with Holographic Mixing} \label{sedfieldholo}
We next want to formulate dynamical equations for the family of
potentials~$A_a$ with~$a=1,\ldots, N$.
A naive idea would be to impose that every potential~$A_a$ should satisfy the
homogeneous Yang-Mills equation
\[ 
D*(F^{A_a} )= 0 \qquad \text{for all~$a \in 1,\ldots, N$}\:. \]
According to~\eqref{JYM}, this equation would describe a nonlinear coupling of each potential~$A_a$ to itself.
However, the resulting set of equations is too simple because we want
that each Yang-Mills potential also interacts with all the other Yang-Mills
potentials. The right procedure for doing so is determined by the
causal action principle and consists of formulating the Yang-Mills equation
directly for the nonlocal potential~$\mathbf{B}$ with holographic 
mixing~\eqref{Bdef}, i.e.\
\beq \label{YMB}
D*(F^{\mathbf{B}}) = 0 \:,
\eeq
where
\beq \label{FYM}
F^{\mathbf{B}} := d\mathbf{B} - i \mathbf{B} \wedge \mathbf{B} \:.
\eeq
Here we can work again with the matrix notation~\eqref{notmatrix} and the
approximation of matrix multiplication~\eqref{Bprodapprox}.

This equation can be justified from the causal action principle, as we now outline. In preparation, we briefly recall how the classical bosonic field equations
can be derived from the causal action principle using the formalism
of the continuum limit (as worked out in~\cite{cfs, nonlocal}).
The starting point is the Dirac equation in the presence of a classical
Yang-Mills potential,
\[ (i \Pdd + \gamma^j A_j^\alpha\, e_\alpha - m) \psi = 0 \:. \]
A-priori, the Yang-Mills potential can be chosen arbitrarily; in particular, it does not
need to satisfy the Yang-Mills equations.
We now consider a distributional solution of this Dirac equation, referred to as the
{\em{kernel of the fermionic projector}}~$P(x,y)$
(for details see~\cite[Sections~2.1.3 and~2.1.4]{fockentangle}
or~\cite[\S1.1.3, \S1.2.4]{cfs}, \cite[Chapter~5]{intro}). Treating the dependence on the Yang-Mills potential perturbatively to second order (for details see~\cite[\S2.1.6]{cfs} 
or~\cite[Chapter~18]{intro})
and analyzing the singularity structure
on the light cone with the so-called {\em{light-cone expansion}}
(a variant of the Hadamard expansion suitable for the perturbation expansion
in Minkowski space; see~\cite{light} or~\cite[Section~2.2]{cfs}, \cite[Chapter~19]{intro}), the resulting formulas of the light-cone expansion
can be inserted into the {\em{EL equations of the causal action principle}}.
Doing so, one finds that the EL equations are satisfied if and only if
the classical Yang-Mills equations hold. In the homogeneous case without
matter fields, these equations take the form
\beq \label{YM}
J^A = 0
\eeq
with~$J^A$ according to~\eqref{JYM}. We note for clarity that 
the light-cone expansion and the continuum limit are gauge covariant.
This guarantees that the resulting field equations are necessarily gauge invariant.
The fact that the current (and not for example the field strength or covariant derivatives of the current) appear in~\eqref{YM} is a consequence of the detailed form
of the causal Lagrangian.

We now move on to the Dirac equation in the presence of the nonlocal potential~$\B$ in~\eqref{dirnonloc}. In this case, the  light-cone expansion can be performed just as for local potentials. One gets exactly the same formulas, with the Yang-Mills potential~$A$
replaced by the nonlocal operator~$\B$.
The only difference is that, when the potentials in the light-cone expansion
are multiplied together, the dephasing effects need to be taken into account.
Describing these effects again with the matrix notation~\eqref{notmatrix},
we again get the homogeneous Yang-Mills equation~\eqref{YM}, but now with the Yang-Mills potential~$A$ replaced by the $N \times N$-matrix potential~$\mathbf{B}$. This gives~\eqref{YMB} and~\eqref{FYM}.

\section{Specifying the Stochastic Background} \label{secbackground}
\subsection{Gauge Freedom with Holographic Mixing}
With~\eqref{dirnonloc} and~\eqref{YMB} we formulated a nonlocal coupled
Dirac-Yang/Mills system. We next analyze the gauge freedom of this system.
At first sight, it is not even clear what ``gauge freedom'' means, because
one could enlarge the usual local gauge transformations by corresponding
nonlocal transformations. A systematic procedure for doing so is to work
again with the matrix notation and the approximation of matrix
multiplication~\eqref{Bprodapprox}.
Given a family of real-valued functions~$\lambda^\alpha_c$
with~$\alpha=1,\ldots, A$ and~$c=1, \ldots, N$, we transform the holographic
Dirac operator~\eqref{dirhol} with potential as in~\eqref{Bdef}
according to
\beq \label{Ugen}
\Dir^\hol \rightarrow \scrU \Dir^\hol \scrU^{-1} \:,\qquad
\scrU = \exp \Big( i \lambda^\alpha_c \starD \mathbf{L}^c \: e_\alpha \Big) \:.
\eeq
Let us work out how the Dirac operator transforms on the linearized level
(the nonlinear corrections will be treated in Section~\ref{secgaugefix}).
Using that the vacuum Dirac operator commutes with the
convolution operators and with the generators of the Yang-Mills field, we obtain
\begin{align*}
\scrU \Dir^\hol \scrU^{-1} &= i \Pdd + \mathbf{B} +
\Big[ i \Pdd, \: \big(-i \lambda^\alpha_c \starD \mathbf{L}^c \: e_\alpha \big) \Big] 
+ \O \big(\lambda \mathbf{B} \big) + \O \big(\lambda^2 \big) \\
&= i \Pdd + \mathbf{B} + (\Pdd \lambda^\alpha_c) \starD \mathbf{L}^c \: e_\alpha
+ \O \big(\lambda \mathbf{B} \big) + \O \big(\lambda^2 \big) \\
&= i \Pdd + \tilde{\mathbf{B}} + \O \big(\lambda \mathbf{B} \big) + \O \big(\lambda^2 \big) \:,
\end{align*}
where the new potential takes the form
\[ \tilde{\mathbf{B}} = \sum_{c=1}^N \gamma^j \big( A^\alpha_{jc} + \partial_j \lambda^\alpha_{c} 
\big) \starD \mathbf{L}^c \: e_\alpha \:. \]
In this way, we can perform a standard $U(1)$ gauge transformation to
each potential~$A^\alpha_c$ separately,
\beq \label{Lorenza}
A^\alpha_{jc}  \rightarrow 
\tilde{A}^\alpha_{jc} = A^\alpha_{jc} + \partial_j \lambda^\alpha_{c} 
+ \O \big(\lambda^2 \big)\:.
\eeq
In view of this transformation law, $\scrU$ describes a gauge
transformation corresponding to the local gauge group~$\U(1)^N$. With this in mind,
we refer to~$\scrU$ as a {\em{generalized gauge transformation}}.

\subsection{Gauge Fixing} \label{secgaugefix}
In order to obtain a well-posed Cauchy problem, we need to remove the
generalized gauge freedom~\eqref{Ugen}.
In the case of abelian gauge fields, the 
standard method for treating the gauge freedom is to fix the gauge
(like for example working in the Lorenz gauge, the temporal gauge or the Coulomb gauge).
Gauge fixing has also been used for non-abelian gauge fields
(for example in the temporal gauge in Minkowski space in~\cite{eardley-moncrief, eardley-moncrief2} and in curved spacetime~\cite{chrusciel-shatah, ghanem}). Alternatively, the Coulomb gauge is considered in~\cite{cronstrom}.
For our purposes, having a perturbative description in mind,
it will be sufficient to use the following simpler procedure.

Using the generalized gauge freedom,
according to~\eqref{Lorenza} we can arrange 
exactly as in classical electrodynamics that
each potential satisfies the Lorenz condition
\[ \partial^j (\tilde{A}^\alpha_c)_j = 0 \:. \]
To be more precise, \eqref{Lorenza} allows us to arrange the Lorenz condition
on the linearized level. But, using the following iteration argument, we can arrange that the
Lorenz condition holds to every order in perturbation theory: We expand~$\tilde{A}$
and~$\lambda$ in powers of the coupling constant,
\[ \tilde{A} = \sum_{p=0}^\infty \tilde{A}^{(p)}\:, \qquad \lambda = \sum_{p=0}^\infty \lambda^{(p)} \:, \]
where for notational simplicity the indices~$\alpha$ and~$c$ were omitted.
Evaluating the Lorenz condition to the order~$p$ gives
\beq \label{lorenzp}
\partial^j (\tilde{A}^{(p)})_j = I \big(\tilde{A}^{(1)}, \cdots, \tilde{A}^{(p-1)} \big) \:,
\eeq
where the right side depends only on lower orders which were already determined
in the previous induction steps. Therefore, in view of~\eqref{Lorenza}
we can satisfy~\eqref{lorenzp} by choosing~$\lambda^{(p)}$ as a solution
of the inhomogeneous wave equation
\beq \label{gaugegen}
\Box \lambda^{(p)} = -I \big(\tilde{A}^{(1)}, \cdots, \tilde{A}^{(p-1)} \big) \:.
\eeq

For the potential~$\mathbf{B}$ in~\eqref{Bdef}, the obtained Lorenz condition
can be expressed as
\beq \label{Lorenz}
\tilde{\mathbf{B}} = \gamma^j \tilde{\mathbf{B}}_j \qquad \text{and} \qquad \partial^j
\tilde{\mathbf{B}}_j = 0 \:.
\eeq
We point out that, with~\eqref{Lorenz},
we have arranged the Lorenz condition as an operator equation on~$\C^N$.
This is quite different from the findings in quantum electrodynamics, where
the Lorenz condition {\em{cannot}} be realized as an operator
equation (see for example~\cite[Section~3-2-1]{itzykson/zuber}). These findings are not a contradiction if one keeps
in mind that the operator~$\mathbf{B}$ acts on the {\em{finite-dimensional}} Hilbert space~$\C^N$.

We now return to the Yang-Mills equation for~$\mathbf{B}$ in the form~\eqref{YMB}
and~\eqref{FYM}. After our generalized gauge transformation which arranges
the Lorenz condition, this equation has become {\em{normally hyperbolic}}. Indeed,
\begin{align*}
\scrU \Dir^\hol \scrU^{-1} &= i \Pdd + \gamma^j \tilde{\mathbf{B}}_j \\
\tilde{\mathbf{B}}_j &= \scrU \mathbf{B}_j \scrU^{-1} + i \scrU \big(\partial_j \scrU^{-1} \big) \\
\mathbf{B}_j &= \scrU^{-1} \tilde{\mathbf{B}}_j \scrU - i \scrU^{-1}
\scrU \big(\partial_j \scrU^{-1} \big) \scrU \\
&= \scrU^{-1} \tilde{\mathbf{B}}_j \scrU - i \big(\partial_j \scrU^{-1} \big) \scrU
= \scrU^{-1} \tilde{\mathbf{B}}_j \scrU + i \scrU^{-1} \big(\partial_j \scrU \big) \:.
\end{align*}
We arranged that
\[ \partial^j \tilde{\mathbf{B}}_j = 0 \:. \]
Thus
\begin{align*}
*(d*d \tilde{\mathbf{B}}) &= \partial_{jk} \mathbf{B}^k - \Box \mathbf{B}_j \\
&=  \partial_{jk}  \big( \scrU^{-1} \tilde{\mathbf{B}}^k \scrU \big)
- \Box \big( \scrU^{-1} \tilde{\mathbf{B}}_j \scrU \big) + \text{(l.o.t.)} \\
&= \scrU^{-1} \big( \partial_{jk} \tilde{\mathbf{B}}^k - \Box \tilde{\mathbf{B}}_j \big) \scrU
+ \text{(l.o.t.)} \\
&= - \scrU^{-1} \big( \Box \tilde{\mathbf{B}}_j \big) \scrU + \text{(l.o.t.)} \:,
\end{align*}
where ``(l.o.t.)'' refers to all lower order terms (i.e.\ contributions involving
at most first derivatives of~$\mathbf{B}$).
In this way, we computed the first summand on the right of
the Yang-Mills equation~\eqref{JYM}. Since all the other summands
are also of lower order (again in the sense that they involve at most first derivatives
of~${\mathbf{B}}$), we conclude that
\beq \label{boxb}
-\Box \tilde{\mathbf{B}}_k + \text{(l.o.t.)} = \scrU J_k \scrU^{-1} \:,
\eeq
where~$J$ is a source term. For this equation, the Cauchy problem
is well-posed. Once~$\tilde{\mathbf{B}}$ is known, we could transform back
to also obtain~$\mathbf{B}$.
In what follows, we shall work in the Lorenz gauge.
For notational simplicity, the tilde will be omitted.

For clarity, we point out that this procedure of solving the Cauchy problem can be
performed order by order in perturbation theory. More precisely, to the order~$p$, we first
arrange the Lorenz gauge using the generalized gauge transformation~\eqref{gaugegen}.
Then we determine~$\tilde{\mathbf{B}}^{(p)}_k$  by solving the Cauchy problem
for the normally hyperbolic equation~\eqref{boxb}.

We finally rewrite the Yang-Mills equation~\eqref{boxb} in the Hamiltonian
form. To this end, we need to rewrite the equation as a first order system in time. For our purposes, it suffices to consider the homogeneous equations.
Setting
\[ 
\Phi = \begin{pmatrix} \mathbf{B} \\ i \partial_t \mathbf{B} \end{pmatrix} \:, \]
we write
\beq \label{hamilton}
i \partial_t \Phi = H_0 \Phi + H^\text{int}_\B(\Phi)
\eeq
with
\[ H_0 = \begin{pmatrix} 0 & 1 \\ -\Delta_{\R^3} & 0 \end{pmatrix} \qquad \text{and} \qquad
H^\text{int}_\B(\Phi) = \begin{pmatrix} 0 \\ - \text{(l.o.t)} \end{pmatrix}
\:. \]
We note that the two components correspond to the usual
canonical field and momentum operators, However, one should keep in mind
that the Hamiltonian equation~\eqref{hamilton} is nonlinear in~$\Phi$,
because the lower order terms in~$H^\text{int}_\B(\Phi)$ include the quadratic
terms in the potential of the Yang-Mills equation as displayed in~\eqref{JYM}.

\subsection{Specifying the Covariance} \label{seccov}
We treat~$\mathbf{B}$ as a stochastic background potential.
We specify the covariance at fixed time by
\beq \label{BGauss}
\bbra A^j_a(t, \vec{x}) \kket = 0 \:,\qquad \bbra A^j_a(t, \vec{x}) \, A^k_b(t, \vec{y}) \kket = h_{a,b}^{jk}(\vec{y}-\vec{x})
\eeq
with covariance matrices~$h^{jk}_{a,b}(y-x)$ which are real and symmetric, i.e.\
\[ 
\overline{h^{jk}_{a,b}(\vec{\xi})} = h^{jk}_{a,b}(\vec{\xi}) = h^{kj}_{b,a}(-\vec{\xi})  \quad \text{for all~$\vec{\xi} \in \R^3$} \:. \]
Similar to~\cite[Theorem~4.1]{fockdynamics} we have the following result.
\begin{Thm} The covariances in~\eqref{BGauss} as well
as the matrix-valued operators~$\mathbf{L}^c$ in~\eqref{notmatrix}
can be chosen in such a way that the potentials~$\mathbf{B}^j$ satisfy the CCR as
operator equations in the sense that
\beq \label{CCR}
\begin{split}
\bbra \big[ & \mathbf{B}^j(t, \vec{x}), \partial_t \mathbf{B}^k(t,\vec{y})  \big] \otimes \mathbf{B}^l(z) \otimes \mathbf{B}^{l'}(z') \kket \\
&\approx i \delta^3(\vec{y}-\vec{x})\: g^{jk} \; \bbra \mathbf{B}^l(z) \otimes \mathbf{B}^{l'}(z') \kket \:,
\end{split}
\eeq
where ``$\approx$'' stands for relative errors with the scaling
\[ \;\times\;\Big( 1 + \O \Big( \frac{\ell_\Lambda}{\ell_\macro} \Big) + \O \Big( \frac{\ell_{\min}}{\ell_\Lambda}\Big) \Big) \]
as well as the UV regularization on the scale~$\varepsilon$.
\end{Thm} \noindent
The proof follows exactly as worked out in~\cite[Sections~4.4 and~4.5]{fockdynamics}:
Disregarding the error terms, the CCR can be arranged by analyzing the covariances
in the commutators and making use of the freedom in choosing the covariances.
The scaling of the error terms can be determined by a stationary phase analysis.

The above choice of the Gaussian covariances can be motivated by Lorentz covariance 
and translations invariance of the effective quantum field theory.
We here prefer to state the CCR in the familiar form in
position space; this is readily obtained from
the result in~\cite[Theorem~4.1]{fockdynamics} by taking the Fourier transform
in~$(t,\vec{x})$ and~$(t', \vec{y})$.
The errors are specified in~\cite[Theorems~4.3 and~5.1]{fockdynamics}.
The reason why, in contrast to the procedure in~\cite{fockdynamics}, we here 
prefer to state the commutation relations only at equal times~$t=t'$ is that the corresponding two-point distribution will now satisfy the
non-linear Yang-Mills equation and can therefore not be stated in closed form.

At this stage, the problem arises that the covariance in the
commutation relations~\eqref{CCR} is not consistent with the
Lorenz condition~\eqref{Lorenz}. This is a well-known problem which
can be resolved for abelian gauge field with the Gupta-Bleuler method
and for non-abelian gauge field with the BRST formalism.
We will discuss these issues in detail in Section~\ref{secBRST}.
For the moment, we can remove the mathematical inconsistencies by
restricting attention to gauge-invariant observables.
Since the nonlinearity will be treated perturbatively
(see Section~\ref{secfeynman} below),
we may here restrict attention to {\em{linear}} fields.
Then, following~\cite{dappiaggi, dappiaggi+siemssen},
the gauge-invariant observables are obtained
by not considering the bosonic field operators
pointwise, but only smeared with closed forms, i.e.\
\beq \label{Bf}
\mathbf{B}(\varphi) := \int_M \mathbf{B} \wedge \varphi
= \int_M \epsilon^{ijkl}\: \mathbf{B}_i(x)\: \varphi_{jkl}\: d^4x \:,
\eeq
where~$\varphi$ is a smooth and compactly supported three-form which is closed, i.e.
\[ d\varphi = 0 \:. \]
Indeed, using the Poincar\'e lemma, $\varphi$ can be written as~$\varphi = dF$
with a two-form~$F$ and thus, applying the Stokes' theorem,
\[ \mathbf{B}(\varphi) = \int_M \mathbf{B} \wedge dF = \int_M d \mathbf{B} \wedge F \:, \]
showing that~$\mathbf{B}(\varphi)$ is indeed gauge independent.

Note that this algebra is too small for performing the perturbation
expansion. A possible way out is to enlarge the space of allowed test
functions (as is done in~\cite{bleuler} in connection with the Gupta-Bleuler method).
We will discuss these issues in detail in Section~\ref{secBRST}.
Before doing so, we now proceed with the formal perturbation expansion.

\section{The Dyson Series} \label{secdyson}

\subsection{Description with Fermionic Fock Spaces}
Clearly, the Dirac equation~\eqref{dirnonloc} describes the dynamics
on the one-particle Hilbert space. However, taking into account that the kernel of the fermionic
projector is composed of many one-particle states, which include the states of negative
energy which form the Dirac sea, the fermions can be described equivalently in the
many-particle picture with fermionic field operators acting on fermionic Fock spaces.
Since this procedure is independent of whether non-abelian or abelian gauge fields
are present, we do not repeat the construction but refer the reader for the details
to~\cite[Section~6.2]{fockdynamics}. We introduce fermionic field operators
by their canonical anti-commutation relations (CARs), which for consistency with~\eqref{CCR}
we write in the equal time formalism as
\[ 
\begin{split}
\{ \Psi^\alpha(t, \vec{x}), \Psi^\beta(t, \vec{y})^\dagger \} &=
\delta^3(\vec{y} - \vec{x})\: \delta_{\alpha \beta} \\
\{ \Psi^\alpha(t, \vec{x}), \Psi^\beta(t, \vec{y}) \} =&\; 0 =
\{ \Psi^\alpha(t, \vec{x})^\dagger, \Psi^\beta(t, \vec{y})^\dagger \} \:.
\end{split} \]
Working in the interaction picture, the interacting Hamiltonian takes the form
\beq H^\text{int}_\Psi = - \int_{\R^3}
\Psi(t, \vec{x})^\dagger \, \alpha^j \mathbf{B}_j(t,x)\,\Psi(t, \vec{x})\: d^3x \:. \label{Hint}
\eeq
One should keep in mind that the potentials~$\mathbf{B}_j$ must satisfy the
homogeneous Yang-Mills equations~\eqref{YMB} and~\eqref{FYM}.

\subsection{The Time-Ordered Product and Feynman Diagrams} \label{secfeynman}
In~\eqref{Hint} the interaction picture was already chosen for the fermionic fields.
Choosing the interaction picture also for the bosonic fields, the
free Hamiltonian~$H_0$ in~\eqref{hamilton} disappears. Thus the dynamics is described by the Schr\"odinger equation for the interacting Hamiltonians in~\eqref{hamilton} and~\eqref{Hint},
\[ i \partial_t \psi = H^\text{int} \psi \qquad \text{with} \qquad
H^\text{int} = H^\text{int}_\Psi + H^\text{int}_\B \:. \]
Here~$\psi$ is a vector in the tensor product of the fermionic and bosonic
Fock spaces.
This equation can be solved with the time-ordered exponential
\[ \psi(t) = \Texp \bigg( -i \int_{t_0}^t e^{-i (t-\tau) H^\text{int}} \:d\tau \bigg) \psi(t_0) \:. \]
In a perturbative expansion, we obtain the usual Dyson series
\begin{align*}
\psi(t) &=  \psi(t_0) -i \int_{t_0}^t d\tau \,H^\text{int}(\tau)\: \psi(t_0) \\
&\quad\:- \int_{t_0}^t d\tau_1 \,H^\text{int}(\tau_1) \int_{t_0}^{\tau_1} d\tau_2 \,H^\text{int}(\tau_2)\: \psi(t_0) + \cdots \:. 
\end{align*}
The standard {\em{Feynman diagrams}} are obtained by taking vacuum expectation
values. Indeed, decomposing the field operators into creation and
annihilation operators, iteratively commuting respectively anti-commuting the
annihilation operators to the right, one gets all the Feynman diagrams of
quantum field theory. The combinatorics of the perturbation expansion as well
as the analytic form of all diagrams (for example rewritten in the usual form as
momentum integrals) are exactly the same as in textbook pQFT, with the only difference
that all diagrams involve an UV regularization on the length scale~$\varepsilon$
(see~\eqref{ellmindef}).
We note for clarity that the boson-boson interaction comes into play as a consequence
of the fact that the Yang-Mills equation in the Hamiltonian form~\eqref{hamilton}
is nonlinear in the bosonic field operators.
As a consequence of the time ordering, all propagators in the expansion
are Feynman propagators.

\section{Representations of the Bosonic Field Operators} \label{secBRST}
In the previous section, we recovered the usual perturbation series of
perturbative quantum field theory (pQFT)
on a formal level. We now explain how to give this formal series a
mathematical meaning. We follow the standard path taken in pQFT as far
as possible and point out the differences.
We first point out that, so far, we deliberately avoided talking about representations of the bosonic field operators. Indeed, choosing such representations
cannot be done in a straightforward way. The basic problem can be understood
from the fact that the CCR written for example
in the so-called Feynman gauge as
\beq \label{CCRexact}
\big[ \mathbf{B}^j(t, \vec{x}), \partial_t \mathbf{B}^k(t,\vec{y})  \big]= i \delta^3(\vec{y}-\vec{x})\: g^{jk}
\eeq
are not compatible with the Lorenz condition~\eqref{Lorenz}.
In our description so far, this inconsistency was not present, because
the field operators~$\mathbf{B}(t,\vec{x})$ acted on the {\em{finite-dimensional}}
vector space~$\C^N$ (see~\eqref{notmatrix}), and the canonical commutation
relations were satisfied only approximately (see~\eqref{CCR}). 
However,
in order to get the correspondence to standard pQFT, one needs to consider
the limiting case of an infinite-dimensional bosonic Fock space~$N \rightarrow \infty$
and vanishing error terms in~\eqref{CCR}. In this limiting case, the
equations~\eqref{Lorenz} and~\eqref{CCRexact} become inconsistent.
The method of restricting attention to gauge-invariant observables (as outlined
after~\eqref{Bf}) is not sufficient for treating the non-linear interaction because,
inside the Feynman diagrams,
the bosonic field operators are in general not tested with gauge-invariant
observables. One way to bypass this problem is to enlarge the algebra
of observables, as is done in~\cite{bleuler} in the context of the Gupta-Bleuler
method of electrodynamics. This method resolves the above inconsistency
by realizing the Lorenz condition not as an operator equation, but only
for suitable expectation values. However, the Gupta-Bleuler method
has the major disadvantage that gauge invariance is no longer manifest.
For this reason, it is preferable to treat the gauge freedom systematically
using Fadeev-Popov ghost fields or the BRST formalism.
Having derived the time-ordered operator expansions, the BRST formalism
can be used in our setting just as well.
Details can be found in the standard textbooks~\cite{henneaux} or~\cite[Chapter~15]{weinberg2},
\cite[Chapter~16]{peskin+schroeder}. 

\Thanks{{{\em{Acknowledgments:}}
We are grateful to the ``Universit\"atsstiftung Hans Vielberth'' for support.
We would like to thank the referees for valuable feedback.
N.K.'s research was also supported by the NSERC grant RGPIN~1105490-2025.
M.R. was supported by the City University of Hong Kong (Start-up Grant 7200748 and CityU Strategic Grant 7005839) and by the Hong Kong Research Grants Council (ECS-Grant 21306524 and GRF-Grant 11303326).
C.D.\ is grateful for the support of Indam, in particular that of the Gruppo Nazionale di Fisica Matematica
(GNFM).
%
%
%
%

\bibliographystyle{amsplain}

\begin{thebibliography}{10}

\bibitem{chrusciel-shatah}
P.T. Chru\'sciel and J.~Shatah, \emph{Global existence of solutions of the
  {Y}ang-{M}ills equations on globally hyperbolic four-dimensional {L}orentzian
  manifolds}, Asian J. Math. \textbf{1} (1997), no.~3, 530--548.

\bibitem{cronstrom}
C.~Cronstr\"om, \emph{The generalisation of the {C}oulomb gauge to
  {Y}ang-{M}ills theory},
  \href{https://arxiv.org/abs/hep-th/9810002}{arXiv:hep-th/9810002} (1998).

\bibitem{qftlimit}
C.~Dappiaggi, F.~Finster, N.~Kamran, and M.~Reintjes, \emph{The quantum field
  theory limit of causal fermion systems}, in preparation.

\bibitem{fockdynamics}
\bysame, \emph{Holographic mixing and {F}ock space dynamics of causal fermion
  systems}, \href{https://arxiv.org/abs/2410.18045}{arXiv:2410.18045
  [math-ph]}, Ann. Henri Poincar{\'e} \textbf{27} (2026), no.~5, 1885--1969.

\bibitem{dappiaggi}
C.~Dappiaggi and B.~Lang, \emph{Quantization of {M}axwell's {E}quations on
  {C}urved {B}ackgrounds and {G}eneral {L}ocal {C}ovariance},
  \href{https://arxiv.org/abs/1104.1374}{arXiv:1104.1374v2 [gr-qc]}, Lett.
  Math. Phys. \textbf{101} (2012), no.~3, 265--287.

\bibitem{dappiaggi+siemssen}
C.~Dappiaggi and D.~Siemssen, \emph{Hadamard states for the vector potential on
  asymptotically flat spacetimes},
  \href{https://arxiv.org/abs/1106.5575}{arXiv:1106.5575 [gr-qc]}, Rev. Math.
  Phys. \textbf{25} (2013), no.~1, 1350002, 31.

\bibitem{eardley-moncrief}
D.M. Eardley and V.~Moncrief, \emph{The global existence of
  {Y}ang-{M}ills-{H}iggs fields in {$4$}-dimensional {M}inkowski space. {I}.
  {L}ocal existence and smoothness properties}, Commun. Math. Phys. \textbf{83}
  (1982), no.~2, 171--191.

\bibitem{eardley-moncrief2}
\bysame, \emph{The global existence of {Y}ang-{M}ills-{H}iggs fields in
  {$4$}-dimensional {M}inkowski space. {II}. {C}ompletion of proof}, Commun.
  Math. Phys. \textbf{83} (1982), no.~2, 193--212.

\bibitem{light}
F.~Finster, \emph{Light-cone expansion of the {D}irac sea in the presence of
  chiral and scalar potentials},
  \href{https://arxiv.org/abs/hep-th/9809019}{arXiv:hep-th/9809019}, J. Math.
  Phys. \textbf{41} (2000), no.~10, 6689--6746.

\bibitem{pfp}
\bysame, \emph{The {P}rinciple of the {F}ermionic {P}rojector},
  \href{https://arxiv.org/abs/hep-th/0001048}{hep-th/0001048},
  \href{https://arxiv.org/abs/hep-th/0202059}{hep-th/0202059},
  \href{https://arxiv.org/abs/hep-th/0210121}{hep-th/0210121}, AMS/IP Studies
  in Advanced Mathematics, vol.~35, American Mathematical Society, Providence,
  RI, 2006.

\bibitem{cfs}
\bysame, \emph{The {C}ontinuum {L}imit of {C}ausal {F}ermion {S}ystems},
  \href{https://arxiv.org/abs/1605.04742}{arXiv:1605.04742 [math-ph]},
  Fundamental Theories of Physics, vol. 186, Springer, Cham, 2016.

\bibitem{nonlocal}
\bysame, \emph{Solving the linearized field equations of the causal action
  principle in {M}inkowski space},
  \href{https://arxiv.org/abs/2304.00965}{arXiv:2304.00965 [math-ph]}, Adv.
  Theor. Math. Phys. \textbf{27} (2023), no.~7, 2087--2217.

\bibitem{fockbosonic}
F.~Finster and N.~Kamran, \emph{Complex structures on jet spaces and bosonic
  {F}ock space dynamics for causal variational principles},
  \href{https://arxiv.org/abs/1808.03177}{arXiv:1808.03177 [math-ph]}, Pure
  Appl. Math. Q. \textbf{17} (2021), no.~1, 55--140.

\bibitem{fockfermionic}
\bysame, \emph{Fermionic {F}ock spaces and quantum states for causal fermion
  systems}, \href{https://arxiv.org/abs/2101.10793}{arXiv:2101.10793
  [math-ph]}, Ann. Henri Poincar\'{e} \textbf{23} (2022), no.~4, 1359--1398.

\bibitem{fockentangle}
F.~Finster, N.~Kamran, and M.~Reintjes, \emph{Entangled quantum states of
  causal fermion systems and unitary group integrals},
  \href{https://arxiv.org/abs/2207.13157}{arXiv:2207.13157 [math-ph]}, Adv.
  Theor. Math. Phys. \textbf{27} (2023), no.~5, 1463--1589.

\bibitem{intro}
F.~Finster, S.~Kindermann, and J.-H. Treude, \emph{{C}ausal {F}ermion
  {S}ystems: {A}n {I}ntroduction to {F}undamental {S}tructures, {M}ethods and
  {A}pplications}, \href{https://arxiv.org/abs/2411.06450}{arXiv:2411.06450
  [math-ph]}, Cambridge Monographs on Mathematical Physics, Cambridge
  University Press, Cambridge, 2026.

\bibitem{bleuler}
F.~Finster and A.~Strohmaier, \emph{{G}upta-{B}leuler quantization of the
  {M}axwell field in globally hyperbolic space-times},
  \href{https://arxiv.org/abs/1307.1632}{arXiv:1307.1632 [math-ph]}, Ann. Henri
  Poincar{\'e} \textbf{16} (2015), no.~8, 1837--1868.

\bibitem{ghanem}
S.~Ghanem, \emph{The global non-blow-up of the {Y}ang-{M}ills curvature on
  curved space-times}, \href{https://arxiv.org/abs/1312.5476}{arXiv:1312.5476
  [math.AP]}, J. Hyperbolic Differ. Equ. \textbf{13} (2016), no.~3, 603--631.

\bibitem{henneaux}
M.~Henneaux and C.~Teitelboim, \emph{Quantization of {G}auge {S}ystems},
  Princeton University Press, Princeton, NJ, 1992.

\bibitem{itzykson/zuber}
C.~Itzykson and J.C. Zuber, \emph{Quantum {F}ield {T}heory}, McGraw-Hill
  International Book Co., New York, 1980, International Series in Pure and
  Applied Physics.

\bibitem{peskin+schroeder}
M.E. Peskin and D.V. Schroeder, \emph{An {I}ntroduction to {Q}uantum {F}ield
  {T}heory}, Addison-Wesley Publishing Company Advanced Book Program, Reading,
  MA, 1995.

\bibitem{weinberg2}
S.~Weinberg, \emph{The {Q}uantum {T}heory of {F}ields. {V}ol. {II}}, Cambridge
  University Press, Cambridge, 1996, Modern Applications.

\end{thebibliography}
\providecommand{\bysame}{\leavevmode\hbox to3em{\hrulefill}\thinspace}
\providecommand{\MR}{\relax\ifhmode\unskip\space\fi MR }
\providecommand{\MRhref}[2]{%
  \href{http://www.ams.org/mathscinet-getitem?mr=#1}{#2}
}
\providecommand{\href}[2]{#2}

\end{document}